\documentclass[epj]{svjour}
\usepackage{graphics}
\begin{document}
\title{Resonant charging of Xe clusters in Helium nanodroplets under intense laser fields}

\author{Christian Peltz \and Thomas Fennel
}                     % Do not remove

\institute{Institute of Physics, University of Rostock, D-18051 Rostock, Germany}

\date{Received: date / Revised version: date}

\abstract{
We theoretically investigate the impact of multiple plasmon resonances on the charging of Xe clusters embedded in He nanodroplets under intense pump-probe laser excitation ($\tau=25$\,fs, $I_0=2.5\times10^{14}{\rm W/cm}^2$, $\lambda=800$\,nm).
Our molecular dynamics simulations on Xe$_{309}$He$_{10000}$ and comparison to results for free Xe$_{309}$ give clear evidence for selective resonance heating in the He shell and the Xe cluster, but no corresponding double hump feature in the final Xe charge spectra is found. Though the presence of the He shell substantially increases the maximum charge states, the pump-probe dynamics of the Xe spectra from the embedded system is similar to that of the free species. In strong contrast to that, the predicted electron spectra do show well-separated and pronounced features from highly efficient plasmon assisted electron acceleration for both resonances in the embedded clusters. A detailed analysis of the underlying ionization and recombination dynamics is presented and explains the apparent disaccord between the resonance features in the ion and electron spectra.
}

\maketitle

\section{Introduction}
\label{intro}
Atomic clusters in intense laser fields provide a rich testing ground for exploring strong-field dynamics and collective phenomena in finite many particle systems \cite{SaaJPB,FenRMP}. Under intense near-infrared laser pulses ($\lambda\sim800\,$nm), the transition of gas-phase clusters into well-isolated finite nanoplasmas results in high energy absorption, rapid cluster explosion, and the emission of highly charged ions \cite{SnyPRL96,KoelPRL99,ZamPRA04}, fast electrons \cite{SprPRA03,KumPRA03,FukPRA03,FenPRL07a}, and energetic photons from the vacuum ultraviolet up to the x-ray range \cite{McPNAT94,DitPRL95,ParPRE00,TerPRE01,PriPRA08}. Because of the opportunity to study the underlying ionization, heating, and decay mechanisms in a nearly background-free environment, intense laser-cluster interactions are of high interest also for various other fields ranging from plasma physics to applied laser-matter research.

Strong enhancements in the energy, yield, or charge states of emitted species \cite{SprPRA03,KumPRA03,FukPRA03,DoepAPB00} can be achieved with laser fields that induce resonant collective heating of nanoplasma electrons (Mie plasmon resonance). When present, the plasmon resonance is typically stronger than additional higher-order nonlinear resonance effects \cite{BauPRL06}. However, in rare-gas clusters where a dense nanoplasma is formed rapidly via tunnel ionization and subsequent electron impact avalanche ionization, the high charge density impedes resonant excitation in early stages of the interaction. The situation is similar for most simple metal clusters. Therefore, a certain cluster expansion is required for efficient resonant heating, making the temporal pulse structure to an efficient control knob for the dynamics. Control of the absorption \cite{ZweiPRA99} and the emission characteristics by the pulse structure has been demonstrated with various schemes, e.g. stretched pulses \cite{KoelPRL99,SprPRA03,FukPRA03,DoepAPB00}, pump-probe techniques \cite{ZweiPRA99,SprPRA00,DoepPRL05,DoepPRA06}, and multi-parameter pulse shaping \cite{ZamPRA04,TruongPRA10}. Typically, resonant plasmon excitations lead to the emission of laser aligned energetic electrons \cite{SprPRA03,KumPRA03,FenPRL07a} and highly charged atomic ions with high kinetic energy \cite{FukPRA03,DoepPRA06}.
Whereas the effect of resonance heating in homogenous clusters is relatively well understood~\cite{SaaPRL03}, there is an ongoing debate on how the possibility of multiple plasmon resonances in core-shell systems affects the interaction dynamics and the final emission spectra. Prominent specimens are clusters in He nanodroplets, as the pickup method with helium nanodroplets is well established in the strong-field cluster community \cite{DoepPCCP07,TigPCCP07}.

Recent theoretical work on Xe clusters in He droplets suggests that (i) nanoplasma generation is triggered by ionization of the embedded cluster and (ii) subsequent avalanche ionization rapidly ionizes the He nanomatrix~\cite{MikPRA08,DoepPRL10}. As the explosion of the He shell is fast, resonant conditions can be achieved at earlier times in the shell than in the embedded cluster. The above calculations show two separate resonance features in the energy absorption. One may expect that signatures of these multiple resonances should be imprinted also in the final emission spectra, e.g. in pump probe experiments. So far, however, no indications have been found in the published data to support the hypothesis that two different optimal delays -- one tuned to a resonance of the rapidly exploding He shell and the other to the later resonance of the embedded cluster -- can be used to enhance the emission of highly charged ions. The available data for embedded clusters appear to be qualitatively similar to experiments on free clusters and have thus often been interpreted by neglecting the detailed dynamics of the helium shell \cite{DoepPRL05,DoepPCCP07,DoepEPJD03}. Nonetheless, it is by now well accepted, that a He shell has decisive impacts on the interaction process. A solid understanding of signatures from multiple resonances in core-shell systems is still lacking.

In the current contribution we investigate this problem theoretically within a pump-probe analysis using a classical molecular dynamics (MD) approach. The MD model accounts for tunnel and electron impact ionization of cluster and shell constituents under inclusion of local plasma field effects and considers electron-ion recombination for the calculation of charge state distributions. As free and embedded model systems we consider Xe$_{309}$ and Xe$_{309}$He$_{10000}$ under two linearly polarized 25\,fs NIR pulses (800\,nm) with intensity $2.5\times10^{14}\,{\rm W/cm}^2$. Our systematic analysis of the spatially resolved energy absorption, the final ion charge state distributions, and the electron energy spectra as function of pulse delay reveals the following key results: (i) irrespective of the presence of the He shell, the final Xe charge states are substantially enhanced for a similar pulse delay and only one optimal delay is found in both cases. The amplitude of the enhancement is larger in the presence of the He shell, which can be attributed to more efficient inner ionization by the first pulse, i.e., prior to any resonance, and more efficient recombination due to stronger screening for nonresonant conditions. (ii) The spatially resolved energy absorption as well as the electron emission spectra give clear evidence for the presence of two resonances with a pronounced double hump in the electron spectra as function of pulse delay. Our results thus indicate that the parallel measurement of electron and ion spectra offers a promising avenue to experimentally resolve multiple resonances in embedded clusters and their consequences for the ion spectra.

The remainder of the text is structured as follows. In Sec. \ref{sec:method} we describe the theoretical approach and the numerical
implementation of the MD model. Section \ref{sec:results} discusses the key stages of the interaction dynamics based on a time-dependent analysis of representative simulation runs (Sec. \ref{sec:time_dep_run}) and the pump-probe spectra of the ions and electrons in relation to the spatially resolved absorption analysis (Sec. \ref{sec:pp_scan}). Conclusions are drawn in Sec. \ref{sec:conclusions}.

\section{Methods}
\label{sec:method}
The explicit quantum mechanical treatment of clusters in intense laser fields is not feasible with current computer technology. We therefore employ classical molecular dynamics, which has been used with great success for describing laser-cluster interactions on a microscopic level, see e.g. Refs. \cite{SaaJPB,FenRMP} and references therein. The basic idea behind is that atomic ionization processes, which are strongly governed by quantum mechanics, are described statistically via appropriate rates, whereas the dynamics of the resulting ions and electrons can be treated classically. A classical picture for plasma electrons and ions is in most cases justified for strong excitations resulting from intense laser irradiation. The laser-matter coupling is described in dipole approximation, which requires the clusters to be sufficiently small such that (i) the spatial variation of the laser field and (ii) laser propagation effects in the target can be neglected. Both requirements are met in our scenarios.

\subsection{Molecular Dynamics Approach }
The following discussion summarizes methodical key aspects of the dynamical model. Note that further details can be found elsewhere~\cite{DoepPRL10,FenPRL07b}.

\begin{figure}
\begin{center}
\resizebox{70mm}{!}{\includegraphics{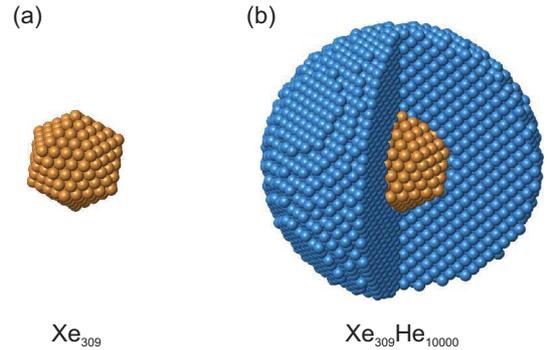}}
\caption{Initial geometry of the investigated free (a) and embedded (b) clusters. In the embedded case, the relaxed icosahedral Xe$_{309}$ is surrounded by a He shell in fcc crystal structure.}
\label{fig:cluster_pic}
\end{center}
\end{figure}

For the simulations we consider Xe$_{309}$ in relaxed icosahedral geometry, see Fig. \ref{fig:cluster_pic}a. A binary Lennard-Jones potential is employed for structural annealing \cite{SchJCP88} and switched off in the simulation of the laser-induced dynamics. The embedded Xe$_{309}$He$_{10000}$ cluster is constructed by cutting a sufficiently large spherical piece from an fcc crystal with the density of liquid helium ($\rho=0.022$\,\AA$^{-3}$, \cite{ToeACH04}). After inserting the Xe$_{309}$ core, He atoms with Xe neighbors closer than the equilibrium He-Xe distance of 4.15\,\AA\, are removed \cite{CheJCP73}, resulting in a compact core-shell compound, cf. Fig. \ref{fig:cluster_pic}b.

Inner ionization of atomic constituents is treated statistically via rates for tunnel ionization (TI) and electron impact ionization (EII). Resulting ions and electrons are then propagated classically in the laser field and under the influence of mutual Coulomb interactions with other plasma particles. The particle trajectories are integrated by a velocity-Verlet propagator with 2.5\,as timestep. The coupled equations of motion read
\begin{equation}
m_i\ddot{\bf r}_i=q_ie{\bf E}_{\rm las}-\nabla_{{\bf r}_i}\sum_{i\neq j}V_{ij},
\end{equation}
where $m_i$, $q_ie$, and ${\bf r}_i$ are the mass, charge, and position of the $i$-th particle and \mbox{${\bf E}_{\rm las}={\bf e}_z E_0 \left[f(t)+f(t-\Delta t)\right]\cos(\omega t)$} describes the electric field of the two linearly polarized laser pulses with temporal delay $\Delta t$, respective peak amplitude \mbox{$E_0=\sqrt{{2I_0}/{c\,\varepsilon_0}}$}, and $\hbar\omega$=1.54\,eV photon energy ($\lambda=$800\,nm). Here $I_0$, $c$ and $\varepsilon_0$ denote the single pulse peak intensity, the vacuum speed of light, and the vacuum permittivity. We employ a gaussian pulse envelope \mbox{$f(t)=\exp(-2\ln2 \,t^2/\tau^2)$}, where $\tau$ is the pulse duration (FWHM). Note, that the single carrier wave results in constructive interference of the pulses for short delays. The pairwise Coulomb interaction $V_{ij}$ is described with a pseudopotential of the form
\begin{equation}
V_{ij}(r_{ij},q_1,q_2)=\frac{e^2}{4\pi \varepsilon_0}\frac{q_i q_j}{r_{ij}} \,{\rm erf}\left(\frac{r_{ij}}{s}\right)
\end{equation}
with the elementary charge $e$, the interparticle distance $r_{ij}$, their charge states $q_i$ and $q_j$, and a numerical smoothing parameter $s$. The latter regularizes the Coulomb interaction and offers a simple route to avoid classical recombination of electrons below the lowest possible quantum mechanical energy level. We use a constant value of $s$ and choose it to be the minimal value, such that the binding energy of an electron to a $q$-fold charged ion is always lower than the $q$-th atomic ionization potential. In our case, the smoothing parameter is determined by Xe and has a value of $s=1.4\,{\rm \AA}$. For the time consuming evaluation of the mutual particle-particle interactions we take advantage of massive parallel computation techniques.

In each timestep, the probability for TI is evaluated from the Ammosov-Delone-Krainov rates \cite{ADK_1986} for each ion, employing the local effective electric field, i.e. the laser field plus the spatially smoothed local plasma field. The plasma field averaging avoids spurious TI events resulting from Rydberg electrons and is realized by evaluation of the pseudopotentials with a smoothing parameter similar to the ground-state Xe-Xe distance. A TI event results in a new plasma electron at the classical tunnel exit and an incremented charge state of the residual ion.

Electron impact ionization is evaluated from effective Lotz cross sections \cite{LotzZP67}, which have been modified to account for local plasma field effects~\cite{FenPRL07b}. Considering an atomic ion in the charged cluster, neighboring ions and electronic screening by quasifree cluster electrons lead to an effective ionization threshold $E_{nl}^*=E_{nl}-\Delta_{\rm env}$ that is lowered with respect to the pure atomic value by the environmental shift $\Delta_{\rm env}$. Here $E_{nl}$ denotes the bare atomic thresholds to remove an electron with principal and angular quantum numbers $n$ and $l$ and $E_{nl}^*$ specifies the corresponding minimal  energy required to lift the electron into the quasi-continuum within the cluster. The shift $\Delta_{\rm env}$ is evaluated directly from the plasma field in the MD simulation using the scheme developed in Ref.~\cite{FenPRL07b}. Therefore the local ionization barrier for a given ionic cell is assumed to be located at half distance to the next neighbor (${\bf r}_{\rm bar}$). The shift $\Delta_{\rm env}$ is then given by the electron potential produced from all charges (including the residual ion) at the point of the barrier (${\bf r}_{\rm bar}$) minus the electron potential in the environmental field alone (residual ion removed) sampled at the position of the ion. Whenever an electron penetrates a new ionic cell, the Lotz cross section $\sigma_{\rm EII}$ is determined including all populated shells of the corresponding ion by employing the effective ionization thresholds $E_{nl}^*$ , the respective shell occupancies, and the kinetic energy of the impinging electron. Ionization is accepted, if the impact parameter $b$ of the impinging electron fulfills the condition $b\le \sqrt{\sigma_{\rm EII}/\pi}$. After a successful EII event the charge state of the corresponding atom or ion is increased and a new plasma electron is generated under the constraint of energy conservation.

\subsection{Spatially resolved energy absorption}
\label{sec:absorption}
\begin{figure}
\begin{center}
\resizebox{0.25\textwidth}{!}{\includegraphics{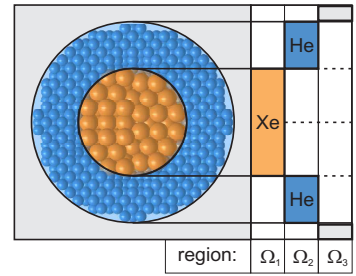}}
\caption{Subdivision of the embedded cluster for the spatially resolved absorption analysis. Regions $\Omega_1$ and $\Omega_2$ contain the Xe core and the He shell, respectively (as indicated). The remainder of simulation space is denoted as the outside region $\Omega_3$. Corresponding radii are updated during cluster expansion. }
\label{fig:absorption_scheme}
\end{center}
\end{figure}

For the identification of resonance heating in different regions of the embedded cluster we perform a spatially resolved analysis of the energy absorbed from the laser field. To this end, the embedded cluster is divided in three different spherical regions, as indicated in Fig. \ref{fig:absorption_scheme}.
The core region $\Omega_1$ contains the Xe cluster and the surrounding region $\Omega_2$ contains the He shell. The rest is denoted as the outside region $\Omega_3$. These regions are time-dependent and the corresponding radii are updated in each timestep. The instantaneous power absorption in region $\Omega_k$ can be expressed as function of the dipole velocity and the laser field by
\begin{equation}
P^{\Omega_k}(t)=\sum_{{\bf r}_i \in \Omega_k} e\,q_i \dot{\bf r}_i \cdot {\bf E}_{\rm las}(t),
\end{equation}
leading to the accumulated energy absorption
\begin{equation}
W_{\rm abs}^{\Omega_k}(t)=\int_{-\infty}^t P^{\Omega_k}(t') dt'.
\end{equation}
The total absorption is then just given by the sum
\begin{equation}
W_{\rm abs}(t)=\sum_{k=1..3}W_{\rm abs}^{\Omega_k}(t) .
\end{equation}

\subsection{Determination of final charge spectra}
\label{sec:det_charge}
To determine the final ion charge spectra, recombination of quasifree electrons with cluster ions after the laser excitation has to be taken into account. Note that the treatment of recombination has strong impacts on the charge distributions and is thus indispensable for predicting realistic spectra~\cite{DoepPRL10,FenPRL07b}. Two relevant mechanisms for electron-ion recombination during cluster expansion are radiative recombination and three-body recombination (TBR). As has been estimated previously~\cite{FenPRL07b}, radiative recombination can be neglected due to low rates~\cite{BetQM77}. TBR, i.e. electron capture after the collision of two quasifree electrons in the vicinity of an ion, proceeds mainly to high Rydberg states of the ion and can therefore be treated classically.
Hence, TBR is automatically included in reasonable approximation within the classical MD propagation. Besides the dominant contribution of TBR, even higher order collisional recombination processes (four-body, five-body, etc.) are accounted for because of the fully microscopic description of classical particle-particle correlations. The key advantage of the direct microscopic treatment of recombination is that no approximations like quasi charge-neutrality or a thermal electron velocity distribution need to be used. In addition, also the local field effects from screening and potentials of neighboring ions are included. 

However, for a qualitative discussion of the recombination dynamics, it is useful to recall basic approximations for the TBR rate from plasma physics. Considering a hydrogen like ion in a high temperature plasma, the TBR rate for electron capture into a bound state with principal quantum number $n$ is given by
$k_n=\alpha \, n^4\,N_e^2/T^2$,
where $N_e$ and $T$ are the number density and temperature of plasma electrons and $\alpha$ is a constant~\cite{HahnPLA97}. Summation over available {\it n}-levels up to the cutoff $n_{\rm c}$ yields, in good approximation, a total TBR rate of
\begin{equation}
k_{\rm tot}=\alpha \ n_{\rm c}^5\,N_e^2/ 5 \,T^2.
\label{eq:k_TBR_tot}
\end{equation}
With the thermal cutoff \mbox{$n_{\rm c}=n_T=(E_1/k_B T)^{1/2}$}, where $E_1$ is the binding energy for \mbox{$n=1$} and $k_B$ is the Boltzmann constant, one finds the well-known $k_{\rm tot} \propto N_e^2 \,T^{-9/2}$ TBR scaling law~\cite{SalzAPHP98}. In contrast to that, when considering a realistic cluster experiment, permanent recombination is assumed only for electron capture in levels with $n<n_{crit}$, where $n_{crit}$ indicates the critical level above which electrons can be re-ionized by the weak ion extraction field of the detector (field ionization cutoff).

When the highly excited cluster expands, the temperature of quasifree electrons decreases rapidly via adiabatic expansion cooling ($T\propto N_e^{2/3}$), leading to a steady decrease of the TBR rate for fixed $n_c=n_{crit}$, cf. Eq.~(\ref{eq:k_TBR_tot}). Hence, relevant $n$-levels with $n<n_{crit}$ are efficiently populated only in early stages of expansion, while the total efficiency of this process is reduced for higher nanoplasma temperature, e.g., for resonant heating. Very high Rydberg levels ($n>n_{crit}$), on the other hand, are populated only at later stages, as high $n$-levels become available only after significant cooling and after a certain degree of expansion (interatomic Coulomb barriers rise with cluster radius). These very high and long-living Rydberg levels are assumed to be re-ionized by the electric field of the ion detector or by other space charge fields.

To measure the result of these complex recombination processes in the MD simulation, we apply the following  simplified scheme. First, as the recombination dynamics is included in the MD, we simply propagate for a sufficiently long time after laser excitation. Electrons are then treated as recombined when localized to an ion (single particle energy below the potential barrier of the respective ionic cell). The inner charge state of an ion is reduced by the number of localized electrons at the time $t=t_{\rm recomb}$. We use a sufficiently large value of $t_{\rm recomb}$, such that the resulting spectra are well-converged. The key assumption behind this simplified treatment is that $n$-levels leading to permanent recombination are populated only in a short time window after laser excitation. As has been checked carefully in a previous study~\cite{FenPRL07b}, recombination processes at longer time scales (for $t\gg t_{\rm recomb}$) produce predominantly very high lying Rydberg electrons that can be re-ionized easily by weak external fields.

\section{Results}
\label{sec:results}
In the following we discuss the simulation results obtained with the above methods for pump-probe excitations of Xe$_{309}$ and Xe$_{309}$He$_{10000}$ by two 25\,fs laser pulses at $I_0=2.5\times10^{14}\,{\rm W/cm}^2$. We begin with a time-dependent analysis of typical runs to highlight the pivotal stages of the interaction processes in the free and embedded clusters and then proceed with the systematic inspection of ion and electron emission spectra.

\subsection{Time-dependent cluster response}
\label{sec:time_dep_run}
\begin{figure*}
\begin{center}
\resizebox{\textwidth}{!}{
  \includegraphics{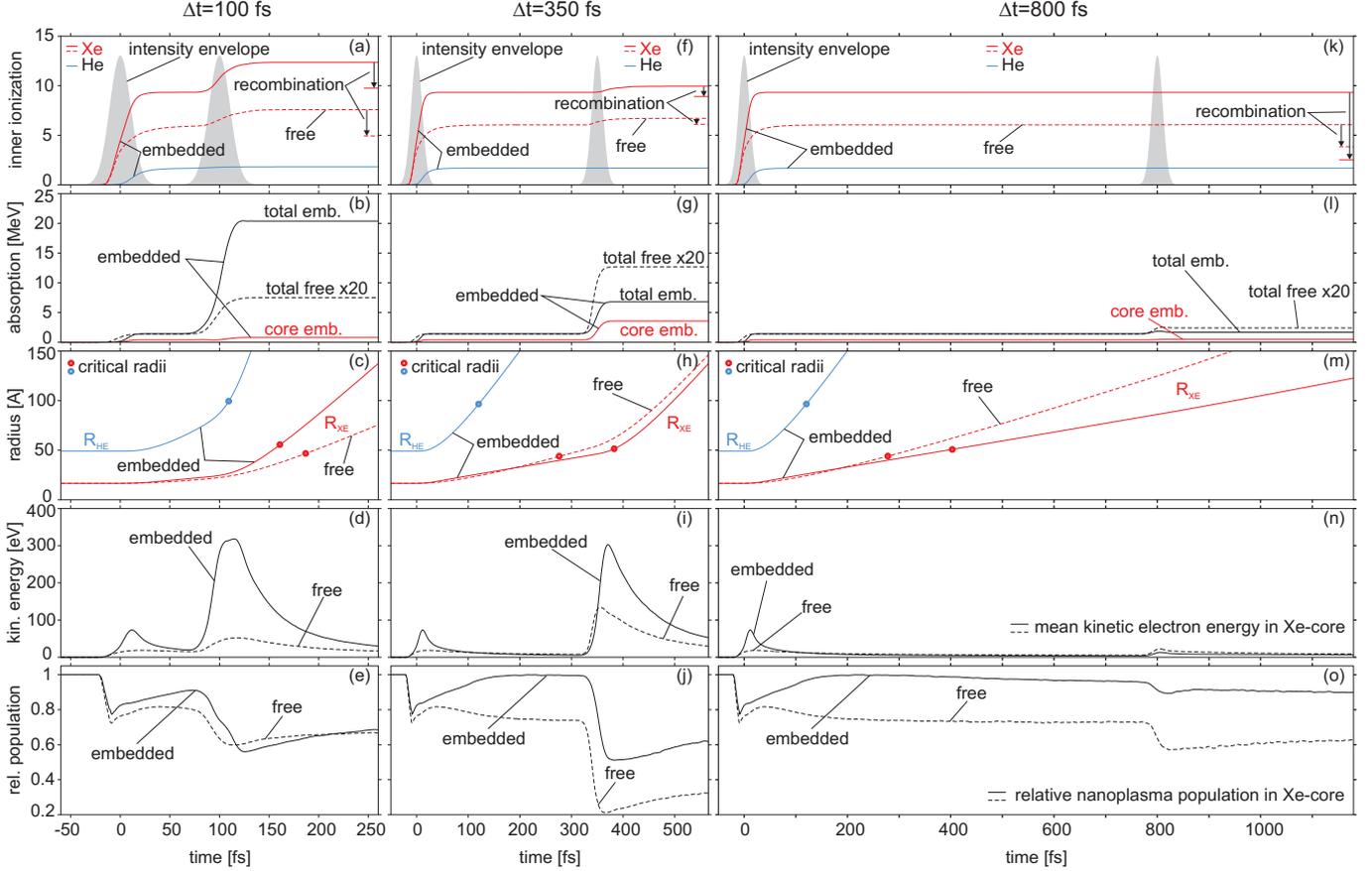}
}
\caption{Calculated evolution of inner ionization, absorption, cluster and matrix radii (circles indicate estimated critical radii for resonant excitation), average Xe-core electron energy, and relative nanoplasma population in the Xe region (top to bottom) for free and embedded Xe$_{309}$ (as indicated) under two 25\,fs laser pulses (delays from left to right: $\Delta t_1 =100\,\mathrm{fs},\Delta t_2 =350\,\mathrm{fs},\Delta t_3 =800\,\mathrm{fs}$) at intensity $I_0=2.5\times 10^{14}\; \mathrm{W/cm^2}$. Note the different scalings of the time axes.}
\label{fig:run_comp}
\end{center}
\end{figure*}

The evolution of selected key observables is displayed in Fig. \ref{fig:run_comp} for three different pulse delays. Two of them are chosen such that resonant heating by the second pulse is induced in the He shell (left panels) or the Xe cluster (middle panels), respectively. The third case is a long delay, where the second pulse cannot excite either of the resonances (right panels).

Focussing on the response to the pump pulse first, our analysis reveals the following picture: Irrespective of the presence of the He shell, ionization begins with tunnel ionization of Xe atoms, see the onsets of inner ionization in Fig. \ref{fig:run_comp}a. Laser heating of the first released electrons induces an impact ionization avalanche that quickly increases the inner Xe ionization levels. Up to the peak of the pump pulse, however, Xe inner ionization evolves very similar in the free and embedded systems, reflecting that the helium does not play a significant role in these very first stages. Subsequently, triggered by the plasma produced in the Xe core, also the He shell becomes rapidly ionized in a second EII avalanche to an average inner charge state of He ions of almost two. In turn, as a feedback effect of the shell activation in the second half of the first pulse, Xe inner ionization is enhanced via EII by almost a factor of two over the free cluster result (compare solid and dashed red curves in Fig.~\ref{fig:run_comp}a). This effect can be attributed to the additional electrons produced in the He shell and the resulting stronger heating. The average kinetic energy in the Xe core region during the first pulse, which is crucial for EII of Xe, reaches an about three times higher peak value in the embedded case, see Fig.~\ref{fig:run_comp}d. After the first pulse, the total energy absorption of the embedded system is about 20 times higher than for the free cluster, cf. dashed and solid black curves in Fig. \ref{fig:run_comp}b.

As a result of the ionization and heating induced by this first excitation step, the Xe cluster and the He shell (if present) begin to expand, see Fig. \ref{fig:run_comp}c. The shell explosion proceeds much more rapidly than that of the cluster due to the low atomic mass of He. Interestingly, the expansion speed of the Xe cluster changes only weakly in the presence of the He shell -- see also Figs.~\ref{fig:run_comp}h and \ref{fig:run_comp}m, where a longer period can be inspected prior to the arrival of the second pulse. Only after several hundred fs a slightly slower Xe expansion is found in the embedded system.

The He shell reaches its critical radius for resonant excitation after about $t=100$\,fs (see Fig. \ref{fig:run_comp}c), which roughly coincidences with the arrival of the second pulse in the $\Delta t=100$\,fs case (left panels). The resonance heating of the He shell results in a total energy absorption that is by more than one magnitude higher than that from the first pulse (Fig. \ref{fig:run_comp}b). Note that the absorption in the core region (red curve in Fig. \ref{fig:run_comp}b) is still fairly low in this case. Nevertheless, though the Xe cluster has not reached its critical radius, absorption is also increased with respect to the first pulse (cf. the free cluster result in Fig. \ref{fig:run_comp}b), but the enhancement is smaller than for the resonantly driven He in the embedded case.
For the intermediate delay of $\Delta t=350$\,fs (middle panels), the Xe cluster (free and embedded) is near the critical radius at the arrival of the second pulse, see Fig. \ref{fig:run_comp}h. This is reflected in a much stronger absorption enhancement in the core region for the embedded system and in the total absorption of the free cluster (cf. Fig. \ref{fig:run_comp}g). Core absorption now exceeds 50\% of the total energy absorption in the embedded case, reflecting efficient and selective resonance excitation of the Xe cluster. In contrast to that, only very weak heating by the second pulse is found for both cases (free and embedded) with the long delay of $\Delta t=800$\,fs (Fig. \ref{fig:run_comp}n), reflecting that no resonance can be excited.

For all scenarios, Xe inner ionization stages are further increased by the second pulse, see Figs.~\ref{fig:run_comp}a, \ref{fig:run_comp}f, and \ref{fig:run_comp}k, but the additional charging is always smaller than the gain during the pump pulse. This behavior reflects that the EII cross sections rapidly decline with increasing charge states.
Further, the additional inner ionization decreases with increasing pulse delay, which reflects that EII is most efficient if the Xe cluster is still dense. Inner ionization becomes less efficient as the Xe cluster expands, as both the flux of impinging electrons as well as the ionization threshold lowering decline. Final ion distributions are calculated by applying the recombination scheme described in Sec. \ref{sec:det_charge} after a typical propagation time of 1 ps after the second pulse. Resulting average final Xe charge states are indicated in Figs. \ref{fig:run_comp}a, \ref{fig:run_comp}f, and \ref{fig:run_comp}k by the vertical arrows.

The weakest charge state reduction by recombination is found with the intermediate pulse delay, where the Xe cluster is resonantly excited. Two effects are responsible for this behavior: First, as electrons in the Xe cluster are driven directly, the electron population in the Xe region is most strongly reduced by the probe pulse. This can be inferred from the evolution of the relative nanoplasma population in the core (see Figs. \ref{fig:run_comp}e, \ref{fig:run_comp}j and \ref{fig:run_comp}o), which gives the ratio of electrons to total ion background charges in the Xe region. A lower population in the expansion phase reduces the recombination because of the density dependence of TBR. Second, the resonant heating strongly increases the electron temperature. The latter is measured by the mean kinetic electron energy in the Xe region, cf. Figs \ref{fig:run_comp}d, \ref{fig:run_comp}i and \ref{fig:run_comp}n. Comparing these average energies in the expanding cluster for a given Xe radius shows that the highest electron temperatures are achieved with $\Delta t=350$\,fs for the free as well as the embedded systems. This trend also suppresses recombination, because of the temperature dependence of the TBR rate, cf. Sec.~\ref{sec:det_charge}.

For both the free and the embedded systems, a slightly larger effect of Xe recombination is found for the short delay of $\Delta t=100$\,fs, where the nanoplasma is less strongly depleted by the probe pulse and also the plasma temperature in the Xe region (for similar Xe radius) is slightly lower than for the $\Delta t=350$\,fs case. It should be emphasized that even the strong resonant heating of the He shell with $\Delta t=100$\,fs is not sufficient to reverse the latter trend. The strongest Xe charge state reduction by recombination occurs with the long delay, where the high nanoplasma population and the low electron temperature in the core lead to efficient TBR. The particularly strong impact of recombination in the embedded cluster is attributed to the weak nanoplasma depletion. The latter effect reflects that the plasma in the surrounding He shell provides cold electrons to efficiently screen the inner Xe core.

\subsection{Pump-probe analysis}
\label{sec:pp_scan}
\begin{figure*}
\begin{center}
\resizebox{0.6\textwidth}{!}{\includegraphics{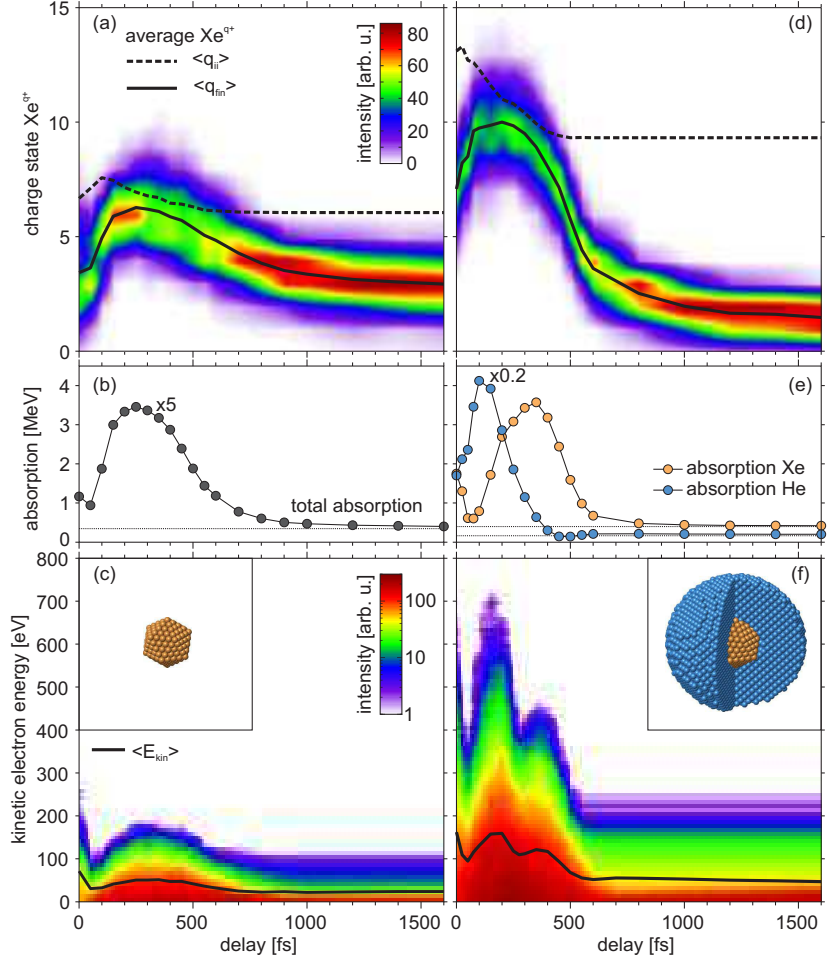}}
\caption{Calculated final Xe charge spectra, energy absorption and electron energy spectra (top to bottom) for pure Xe$_{309}$ (left) and Xe$_{309}$He$_{10000}$ (right) under two 25 fs pulses (intensity: $I_0=2.5\times 10^{14} \mathrm{W/cm^2}$) as function of pulse delay. Average values of inner ionization $\left\langle q_{\mathrm{ii}}\right\rangle$ (dashed)  and final charge states $\left\langle q_{\mathrm{fin}}\right\rangle$ (solid) are given in panel (a) and (d). Solid curves in panels (e) and (f) represent the average kinetic energy $\left\langle E_{\mathrm{kin}}\right\rangle$ of emitted electrons. }
\label{fig:spec_abs_pic}
\end{center}
\end{figure*}

In the next step we provide a more systematic analysis of the pump-probe effects by focussing on a smaller set of observables. Therefore the pulse delay has been scanned in a series of simulations and the resulting ion (upper panels) and electron spectra (lower panels) have been determined, see Fig. \ref{fig:spec_abs_pic}. The final emission spectra for the free (left panels) and embedded clusters (right panels) are shown in comparison to the energy absorption (middle panels).

For both the free and embedded clusters, the final Xe ion distributions exhibit a pronounced pump-probe dynamics (colored data in Figs. \ref{fig:spec_abs_pic}a and \ref{fig:spec_abs_pic}d). The Xe charge states are maximized for similar optimal pulse delays of about $\Delta t\approx200-300$\,fs, producing substantially higher ionization levels than for much smaller or much larger pulse separations. The ion distributions are rather narrow with a typical spread between 3-5 charge states. 

Inspection of the results for free Xe$_{309}$ reveals a pump-probe dynamics of the charge states that resembles the typical signatures of resonance enhancement. More precisely, the optimal pulse delay for highest charging is in good agreement with the peak of the total energy absorption (cf. Fig. \ref{fig:spec_abs_pic}a and \ref{fig:spec_abs_pic}b), reflecting plasmon enhanced ionization. The main reason for the pronounced pump-probe dynamics in the ion spectra is, however, not the variation of inner ionization as function of pulse separation, but the strong effect of the delay on outer ionization and recombination. 
Comparison of the mean values for inner and final Xe ionization in Fig. \ref{fig:spec_abs_pic}a shows, that inner ionization is almost constant ($\langle q_{\rm ii}^{Xe}\rangle\approx 6.5$) for all delays, with only a slight increase towards small delays (dashed line). The major effect for the enhancement at the optimal delay is the weak recombination at resonance, where the average final Xe stages are close to inner ionization (compare dashed and solid curves). A clear marker for optimal resonance excitation of the free cluster with $\Delta t\approx300$\,fs is the strongly enhanced emission of energetic electrons, see the electron spectra in Fig. \ref{fig:spec_abs_pic}c. Note that the feature at time zero can be attributed to constructive interference of the pulses whereas the asymptotic yield for long delays reflects the single pulse result.

Focussing on the ion spectra from embedded clusters, a much stronger pump probe dynamics and still only one optimal delay are observed. At first glance, no clear effect from two resonances can be found in the ion spectrum, though the spatially resolved energy absorption in the He shell and in the Xe core indicate their selective excitations (cf. Fig. \ref{fig:spec_abs_pic}e).
When compared to the free cluster, the maximum final charge states at optimal delay are about 50\% higher and ionization stages with long delays are lower. Inner ionization, though substantially increased over the free cluster case, is again only weakly dependent on the delay (when compared to the amplitude of the final charge states), but with a stronger enhancement at short delays. Note that the two small humps in the average Xe inner ionization at $\Delta t=100$ and 300\,fs  may be interpreted as direct signatures from the shell and cluster resonances (cf. Fig. \ref{fig:spec_abs_pic}d). Nevertheless, the stronger inner ionization increase at small delays is clearly ascribed to more effective heating of the helium that surrounds the still compact Xe core (density effect).

Like in the free cluster case, most of the strong dynamics in the final charge distribution is an effect of delay dependent recombination. The weakest charge state reduction (cf. dashed and solid curve in Fig. \ref{fig:spec_abs_pic}d) is found for a delay of $\Delta t=300$\,fs where the Xe core can be resonantly excited. For long delays, recombination is particularly efficient in the embedded cluster, as the ionized He shell provides additional electrons for efficient screening of the Xe core. The maximum final Xe charge states are found for a delay between the optimal delay values for resonantly exciting the He shell or the Xe core, i.e. near $\Delta t=200$\,fs. This intermediate delay reflects a compromise between inner ionization, which is enhanced towards the He resonance, and weak recombination, which occurs close to the cluster resonance.

Hence, besides a stronger pump probe dynamics and higher peak ionization levels, our model predicts no clear signature from multiple resonances in the final Xe spectra. In contrast to that, there is a pronounced double hump structure in the electron spectra, see Fig \ref{fig:spec_abs_pic}f. The two strong peaks in the maximum energy of emitted electrons near $\Delta t=150$ and 370\,fs can be directly assigned to plasmon enhanced electron acceleration \cite{FenPRL07a} in the He shell and in the Xe core, respectively. Both coincide well with the absorption peaks in the respective regions, cf. Fig. \ref{fig:spec_abs_pic}e.

\section{Conclusions}
\label{sec:conclusions}
We have studied the ionization dynamics of Xe clusters embedded in He nanodroplets under strong-field pump-probe excitation to analyze the impact of multiple plasmon resonances in the core-shell compound.

The key findings are: (i) For the free and the embedded clusters we observe plasmon enhanced charging with pronounced enhancements of the final Xe charge states for similar optimal pump-probe delays. (ii) In the presence of the He shell, Xe ionization stages at optimal pulse delay are higher than for the free cluster, mostly due to more efficient inner ionization by the first pulse. (iii) The spatially resolved energy absorption analysis gives evidence that selective resonant excitation of the shell or the cluster is possible by a shorter or a longer delay, but without a corresponding double-hump feature in the ion spectra. Our analysis supports that the intermediate optimal delay for highest Xe charging reflects that both strong inner ionization and weak recombination are necessary to generate high ionization stages.
(iv) In contrast to that, the two resonances do leave separate fingerprints in the electron spectra, where highly efficient electron acceleration is found for the respective delays of the shell or core resonance.

These results thus indicate that a He resonance, even if present, may not be observed as a separate feature in the ion spectra.
As direct cluster heating still plays a pivotal role, a shell induced inner ionization enhancement only shifts the optimal delay for highest charging of cluster constituents.
Our theory thus offers an explanation for the absence of corresponding features in the pump-probe experiments published so far. The strong features in our calculated electron spectra, however, suggest that experiments with parallel analysis on ion and electron spectra would offer a promising route to identify the presence and time-structure of multiple resonances for clarifying their link to the ion spectra. For such measurements very short pulses would be advantageous.

Financial support by the Deutsche Forschungsgemeinschaft within the SFB 652/2 and computer time provided by the HLRN computing center are gratefully acknowledged.

\end{document}